%% file: mb_arxiv.tex
\documentclass{article}
\usepackage[comma,square,numbers,sort&compress]{natbib}
\usepackage{lineno}
\usepackage{units}
\usepackage{graphicx}  
\usepackage[left=3cm, right=3cm, top=3cm]{geometry}
\usepackage{hyperref}

\makeindex

\include{commands}

\begin{document}

\title{Phenomenology of soft QCD:\\the role of minimum-bias measurements}

\author{Jan Fiete Grosse-Oetringhaus}
\date{\footnotesize CERN, Geneva, Switzerland, jgrosseo@cern.ch}

\maketitle

\begin{abstract}
This chapter summarizes minimum-bias measurements at the Large Hadron Collider. In particular, the pseudorapidity density, the transverse-momentum spectra, the multiplicity distribution, the correlation of average transverse momentum and the multiplicity, and a measurement of minijets are presented. In addition to an overview of the results obtained to date at the LHC, the experimental challenges of defining particle and event sample and correcting to this sample are discussed.
\end{abstract}

\vspace{0.5cm}
\noindent \emph{This document corresponds to Chapter~12 of Ref.~\cite{Bartalini:2017jkk}.}

\label{EXP-MB-LHC}

\section{Introduction}

The study of minimum-bias (MB) physics comprises in its most general terms all signals which can be well extracted experimentally without the use of triggers which enhance certain, more rare, events in the data stream over the typical \emph{average} event. The experimental apparatus should bias the least (hence minimum-bias or zero-bias) the ensemble of events. Contrary to this conceptually simple event selection, the phenomenology of minimum-bias events is very rich: the underlying processes are dominated by QCD in the non-perturbative regime (at small $Q^2$). Fragmentation and hadronization of the partons produced in the collision as well as multiple-parton interactions play an important role. Theoretically it is difficult to describe these mechanisms from first principles (see also Chapter~10 in Ref.~\cite{Bartalini:2017jkk}) which makes the experimental study of these collisions crucial. 

In order to study this regime, ideally the full phase-space distribution of produced particles and their correlations should be measured, which is given by the probability to find a number of particles $N$ of type $i$ and momentum vector $p$:
\begin{equation}
  P_N(p^i_1, p^i_2, ...).
\end{equation}
It is difficult to measure such a complete observable experimentally. Therefore, one starts more modest, by measuring the event-averaged number of produced particles neglecting their type as a function of a single kinematic property (e.g. as a function of the pseudorapidity d$N$/d$\eta$ or the transverse momentum d$N$/d$\pt$). These measures neglect correlations between the produced particles and characterize the average collision. Measuring the probability distribution of the number of produced particles $P(N)$ contains some degree of correlation between the particles. Similarly, the event dynamics is addressed by measuring the mean transverse momentum as a function of multiplicity. Both these observables show a particular sensitivity to MPIs, discussed below, and are of interest to be studied also at higher multiplicity (see Chapter~15 in Ref.~\cite{Bartalini:2017jkk}).
A direct access to the number of parton interactions can be obtained by studying so-called minijets defined as the particles originating from the same $2 \rightarrow 2$ scattering at low $Q^2$ of a few \unit{GeV/c}. At these scales, the number of particles per parton is of the order of 1, and hence traditional jet finding methods are not applicable, instead statistical approaches can be used.

This chapter will first discuss the experimental challenges of minimum-bias measurements. Subsequently, results from the Large Hadron Collider will be presented. Finally, an outlook for the future of these types of measurements is given.

\section{Experimental Challenges}

The LHC detectors are complex multi-million channel devices designed to record traces of particles from a momentum of \unit[100]{MeV/c} to several TeV/c at MHz collision rates. It is not surprising that this complexity actually results in a larger number of experimental challenges than e.g. in historic bubble chamber experiments which had $4\pi$ coverage and very little material.

The detector effects that need to be corrected for stem from the fact that due to the detector material, dead areas in the detector and the efficiency of electronics and algorithms not every particle is reconstructed. Furthermore, interactions with material and decays of instable particles can create additional particles. Due to these inefficiencies, it may also happen that a certain event is not seen by the detector at all, resulting in a miscount of the total number of occurred collisions which also has to be corrected for.

The first non-trivial step is to define for which particles and for which collisions an observable is to be measured. The choices adopted by the LHC experiments are introduced in the following.

\subsection{Primary-particle definition}

The particles which shall be part of the result are called the \emph{primary particles}. Due to the design of the LHC detectors, the measurements discussed in this chapter have been performed for charged particles. Most of the produced particles are instable. Particles like the $\rho$ which decay strongly, decay almost instantaneously and the decay products are therefore included among the primary particles. Weakly decaying particles need a special treatment, e.g. the $K^0$ and $\Lambda$. Those are neutral, thus not part of the charged primary particles, but due to their decay into charged particles they become part of the measured sample. These are referred to as \emph{secondary particles} which need to be corrected for. Further sources of secondary particles are particles produced by interaction of primary particles with the detector material (e.g. $\gamma \rightarrow e^+e^-$).

The experiments at the LHC have adopted similar conventions for primary particles with minor differences. ATLAS~\cite{Aad:2010ac} and LHCb~\cite{Aaij:2011yj} count particles produced in the collision or produced by decays of particles with a proper lifetime $\tau < \unit[30]{ps}$ and \unit[10]{ps}, respectively. As there are no known particles with lifetimes in the range \unit[10--30]{ps}~\cite{Olive:2016xmw}, these definitions are identical.
ALICE~\cite{Aamodt:2010ft} and CMS~\cite{Khachatryan:2010xs} use a definition which includes products of strong and electromagnetic decays, but excludes products of weak decays and hadrons originating from secondary interactions. Later CMS clarified the definition to include decay products of particles with proper lifetimes less than \unit[1]{cm}~\cite{Chatrchyan:2011av}. In practice, these are identical to the definition using the proper lifetime given before. However, CMS measures only hadrons, therefore excluding leptons from their measurement. 
Lately, ATLAS revisited their definition~\cite{Aad:2016mok} and excludes particles with a proper lifetime between \unit[30]{ps} and \unit[300]{ps} from their primary-particle definition. This change removes strange baryons from the sample for which the reconstruction efficiency was found to be very low. These differences in the definitions result in only few percent effects on the integrated yields but can be relevant in certain momentum regions. In direct comparisons, these have to be carefully considered.

\begin{figure}[t]
\centerline{\includegraphics[width=9cm]{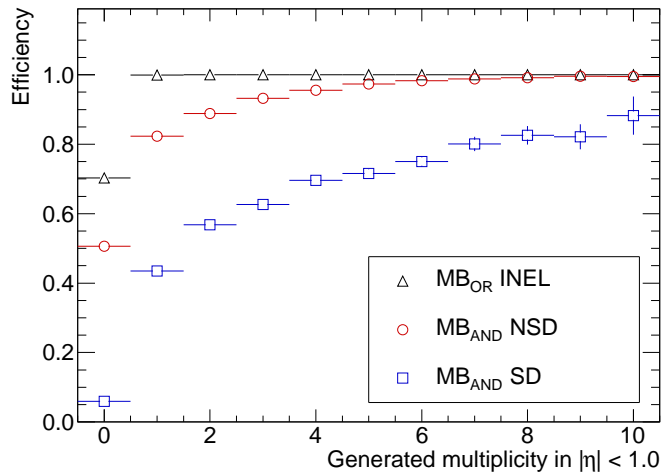}}
\caption{Event selection efficiency of the ALICE experiment at $\s = \unit[0.9]{TeV}$ for a single-arm trigger and inelastic collisions (triangles) and a double-arm trigger for non-single diffractive (circles) and single-diffractive collisions (squares). Figure from Ref.~\cite{Aamodt:2010ft}.}
\label{fig:trigger_efficiency}
\end{figure}

\subsection{Event-Sample Definition}

Inelastic hadronic collisions are divided based on the occurring processes into diffractive and non-diffractive collisions. In diffractive collisions one (or both) of the incoming particles retain their quantum numbers except possibly the spin which are then called single-diffractive (and double diffractive), respectively. Historically, measurements were presented for non-single-diffractive (NSD) collisions excluding the single-diffractive (SD) contribution which was difficult to trigger on. Furthermore, measurements were performed for all inelastic collisions (diffractive + non diffractive) requiring in particular for the SD component significant model-dependent corrections.

Results at the LHC were published for both, NSD and inelastic event classes. In addition, it was realized that the efficiency to measure events with low multiplicity, a region particularly dominated by SD collisions, is rather low, while almost all events with higher multiplicity are measured. Figure~\ref{fig:trigger_efficiency} shows exemplarily the selection efficiency of the ALICE experiment as a function of the number of charged particles $\nch$. It can be seen that as soon as one particle enters the acceptance of the detector, the efficiency to record the event is 100\% when a single-arm trigger is used. Such a trigger is typically used for the measurement of the inelastic event class. For the NSD event class, double-arm coincidence triggers (either during data-taking or in the analysis itself) are employed. These suppress a significant part of the SD contribution as illustrated in Fig.~\ref{fig:trigger_efficiency}.

\begin{table}[th!]
\caption{List of LHC minimum-bias results. In the results column, the letters denote $d\nch/d\eta$ (A), $d\nch/d\pt$ (B), $P(\nch)$ (C), and $\meanpt$ vs. $\nch$ (D). Particle-level event classes are denoted $L^i$ where $i$ is an index; their definitions are given below the table.}
\begin{tabular}{@{}ccccc@{}} \hline
Experiment                      & $\s$ (TeV)     & Events ($\times 10^6$)       & Event classes         & Results \\ \hline
ALICE~\cite{Aamodt:2009aa}       & 0.9            & 0.0003                & INEL, NSD             & A   \\
ALICE~\cite{Aamodt:2010ft}       & 0.9, 2.36      & 0.15, 0.04            & INEL, NSD             & AC   \\
ALICE~\cite{Aamodt:2010pp}       & 0.9, 2.36, 7   & 0.05, 0.04, 0.24      & $L^{\rm a}$        & AC   \\
ALICE~\cite{Aamodt:2010my}       & 0.9            & 0.34                  & INEL, NSD             & B   \\ 
ALICE~\cite{Abelev:2013ala}      & 0.9, 2.76, 7   & 6.8, 65, 150          & INEL                  & B \\
ALICE~\cite{Abelev:2013bla}      & 0.9, 2.76, 7   & 6.8, 65, 150          & INEL                  & D \\
ALICE~\cite{Abelev:2013sqa}      & 0.9, 2.76, 7   & 7, 27, 204            & --                    & Minijets \\
ALICE~\cite{ALICE-PUBLIC-2013-001} & 0.9, 7         & 2.9, 2.7              & $L^{\rm b}$, $L^{\rm c}$, $L^{\rm d}$ & AC \\
ALICE~\cite{Adam:2015gka}        & 0.9, 2.76, 7, 8 & 7.4, 34, 404, 31     & INEL, NSD, $L^{\rm a}$ & AB \\ 
ALICE~\cite{Adam:2015pza}        & 13             & 1.5M                  & INEL, $L^{\rm a}$  & AB \\ \hline
ATLAS~\cite{Aad:2010rd}          & 0.9            & 0.46                  & $L^{\rm e}$        & ABCD \\
ATLAS~\cite{Aad:2010ac}          & 0.9, 2.36, 7   & 0.36, 0.006, 10       & $L^{\rm e}$, $L^{\rm f}$, $L^{\rm g}$, $L^{\rm h}$, $L^{\rm i}$ & ABCD \\
ATLAS~\cite{ATLAS-CONF-2010-101} & 0.9, 7         & 0.36, 10              & $L^{\rm c}$, $L^{\rm d}$ & ABCD \\
ATLAS~\cite{Aad:2015wga}         & 2.76           & 87$^\dagger$          & INEL                  & B \\
ATLAS~\cite{Aad:2016mok}         & 13             & 9                     & $L^{\rm c}$, $L^{\rm e}$        & ABCD$^\ddag$ \\ 
ATLAS~\cite{Aad:2016xww}         & 8              & 9                     & $L^{\rm e}$, $L^{\rm f}$, $L^{\rm g}$, $L^{\rm j}$, $L^{\rm k}$ & ABCD$^\ddag$ \\
ATLAS~\cite{Aaboud:2016itf}      & 13             & 9                     & $L^{\rm f}$        & ABCD$^\ddag$ \\ \hline
CMS~\cite{Khachatryan:2010xs}    & 0.9, 2.36      & 0.07, 0.02            & NSD                   & AB$^\amalg$ \\
CMS~\cite{Khachatryan:2010us}    & 7              & 0.07                  & NSD                   & AB$^\amalg$ \\
CMS~\cite{Khachatryan:2010nk}    & 0.9, 2.36, 7   & 0.25, 0.02, 0.6       & NSD                   & C$^\amalg$ \\
CMS~\cite{CMS-PAS-QCD-10-024}    & 0.9, 7         & 6.1, 0.8              & $L^{\rm c}$, $L^{\rm d}$, $L^{\rm l}$, $L^{\rm m}$ & A \\
CMS~\cite{Chatrchyan:2011av}     & 0.9, 7         & 6.8, 25$^\dagger$     & NSD                   & B$^\amalg$ \\ 
CMS/TOTEM~\cite{Chatrchyan:2014qka} & 8           & 3.4                   & $L^{\rm n}$, $L^{\rm o}$ & A$^\amalg$ \\ 
CMS~\cite{Khachatryan:2015jna}   & 13             & 0.17                  & INEL                  & A$^\amalg$ \\
CMS~\cite{CMS-PAS-FSQ-15-008}    & 13             & 3.9                   & INEL, $L^{\rm l}$, $L^{\rm p}$  & A \\ \hline
LHCb~\cite{Aaij:2011yj}          & 7              & 3                     & $L^{\rm q}$        & AC \\
LHCb~\cite{Aaij:2014pza}         & 7              & 3                     & $L^{\rm r}$        & ABC \\ \hline
\end{tabular}
\begin{minipage}[l]{22.5cm}
\begin{flushleft}
\footnotesize
\vspace{5pt}
$L^{\rm a}$ At least 1 charged particle within $|\eta| < 1$.\\
$L^{\rm b}$ At least 1 charged particle within $|\eta| < 0.8$ and $\pt > \unit[0.15]{GeV/c}$.\\
$L^{\rm c}$ At least 1 charged particle within $|\eta| < 0.8$ and $\pt > \unit[0.5]{GeV/c}$.\\
$L^{\rm d}$ At least 1 charged particle within $|\eta| < 0.8$ and $\pt > \unit[1]{GeV/c}$.\\
$L^{\rm e}$ At least 1 charged particle within $|\eta| < 2.5$ and $\pt > \unit[0.5]{GeV/c}$.\\
$L^{\rm f}$ At least 2 charged particle within $|\eta| < 2.5$ and $\pt > \unit[0.1]{GeV/c}$.\\
$L^{\rm g}$ At least 6 charged particle within $|\eta| < 2.5$ and $\pt > \unit[0.5]{GeV/c}$.\\
$L^{\rm h}$ At least 20 charged particle within $|\eta| < 2.5$ and $\pt > \unit[0.1]{GeV/c}$.\\
$L^{\rm i}$ At least 1 charged particle within $|\eta| < 2.5$ and $\pt > \unit[2.5]{GeV/c}$.\\
$L^{\rm j}$ At least 20 charged particle within $|\eta| < 2.5$ and $\pt > \unit[0.5]{GeV/c}$.\\
$L^{\rm k}$ At least 50 charged particle within $|\eta| < 2.5$ and $\pt > \unit[0.5]{GeV/c}$.\\
$L^{\rm l}$ At least 1 charged particle within $|\eta| < 2.4$ and $\pt > \unit[0.5]{GeV/c}$.\\
$L^{\rm m}$ At least 1 charged particle within $|\eta| < 2.4$ and $\pt > \unit[1.0]{GeV/c}$.\\
$L^{\rm n}$ At least 1 charged particle within $5.3 < |\eta| < 6.5$.\\
$L^{\rm o}$ At least 1 charged particle within each $5.3 < \eta < 6.5$ and $-6.5 < \eta < -5.3$.\\
$L^{\rm p}$ Additional event selections to enhance inelastic and diffractive contributions.\\
$L^{\rm q}$ At least 1 charged particle within $2.0 < \eta < 4.5$.\\
$L^{\rm r}$ At least 1 charged particle within $2.0 < \eta < 4.8$, $\pt > \unit[0.2]{GeV/c}$, $p > \unit[2]{GeV/c}$. \\
$^\dagger$ Includes events enhanced by a high \pt trigger. \\
$^\ddag$ Primary-particle definition excludes strange baryons. \\
$^\amalg$ Primary-particle definition excludes leptons.
\end{flushleft}
\end{minipage}

\label{table:mb_papers}
\end{table}

As the corrections for events which are not triggered upon are model-dependent, so-called \emph{particle level} event classes have been added. These refer to events which have at least a certain number of particles within the detector acceptance (e.g. in $|\eta| < 2.5$ and $\pt > \unit[0.5]{GeV/c}$). These classes reduce the model dependence of the measurement significantly and thus the related uncertainties. However, they can only be compared to models which have a Monte Carlo implementation because the event selection has to be reproduced in the model.

\subsection{Corrections}

The measured distributions are corrected for tracking efficiencies and acceptance as well as contamination by secondary particles. In addition, the finite trigger and selection efficiencies are corrected for. These effects lead to a smearing of the measured quantities. Therefore, for certain observables like the $P(\nch)$ where the spectrum is steeply falling, the measured distribution needs to be unfolded. While mathematically unfolding is an ill-posed problems, regularized unfolding which makes  assumptions on the unfolded spectrum allows to recover the underlying distribution. A discussion of these methods can be found in Refs.~\cite{Blobel:1984ku,Blobel:2002pu,GrosseOetringhaus:2009kz}.

\section{Results}

Despite the fact that the LHC is built as a discovery machine reaching very large integrated luminosities, it is also optimally suited for minimum-bias physics. The high-precision detectors as well as the versatility of the LHC where special runs with low instantaneous luminosity and special beam configurations are possible gave rise to an unprecedented data sample. While the first measurement was done with just 284 events~\cite{Aamodt:2009aa}, later measurements used hundreds of millions of events~\cite{Abelev:2013ala}. Table~\ref{table:mb_papers} lists all relevant papers of the LHC experiment addressing the minimum-bias distributions discussed in this chapter. The precision and the reach in $\pt$ and multiplicity are unprecedented and the wealth of results is a legacy enabling the theory community to develop models with an accurate description of the non-perturbative QCD components which ultimately constitutes the bulk of the particles produced in LHC collisions. The following sections are only able to present a subset of these results with the aim of illustrating the different observables, what was learned and their potential.

\begin{figure}[t]
\centerline{\includegraphics[width=8cm]{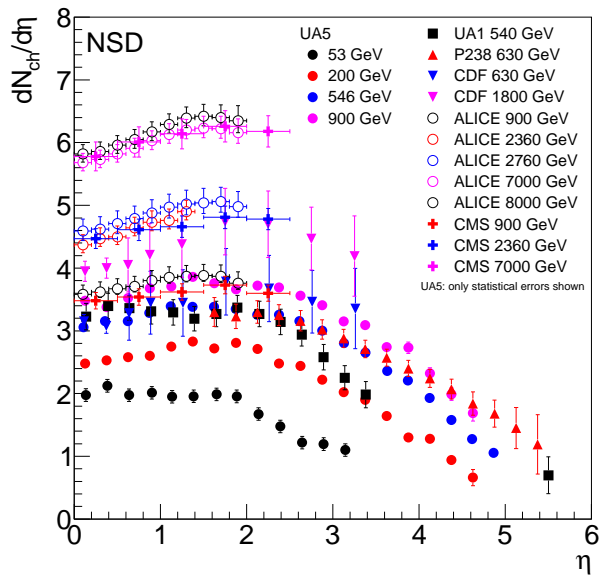}}
\caption{The pseudorapidity density $d\nch/d\eta$ for NSD collisions over more than two orders of magnitude in $\s$. Data from Refs.~\cite{Arnison:1982rm, Alner:1985zc, Ansorge:1988kn, Abe:1989td, Harr:1997sa, Aamodt:2010ft, Adam:2015gka, Khachatryan:2010xs, Khachatryan:2010us}. As this figure presents only results for NSD collisions, no data for \unit[13]{TeV} and from ATLAS and LHCb are shown.}
\label{fig:mb_dndeta}
\end{figure}

\begin{figure}[t]
\centerline{\includegraphics[width=8cm]{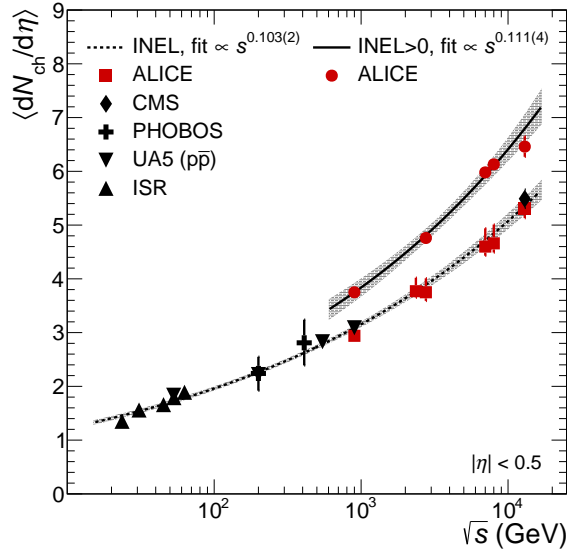}}
\caption{The pseudorapidity density $d\nch/d\eta$ at $\eta = 0$ as a function of $\s$ for inelastic collisions and collisions with at least one particle within $|\eta| < 1$. Figure from Ref.~\cite{Adam:2015pza}.}
\label{fig:mb_dndeta_vs_sqrts}
\end{figure}

\subsection[{Pseudorapidity density $d\nch/d\eta$ and transverse-momentum spectra $d\nch/d\pt$}]{Pseudorapidity density $d\nch/d\eta$ and \newline transverse-momentum spectra $d\nch/d\pt$}

\begin{figure}[t!]
\centerline{\includegraphics[width=8cm]{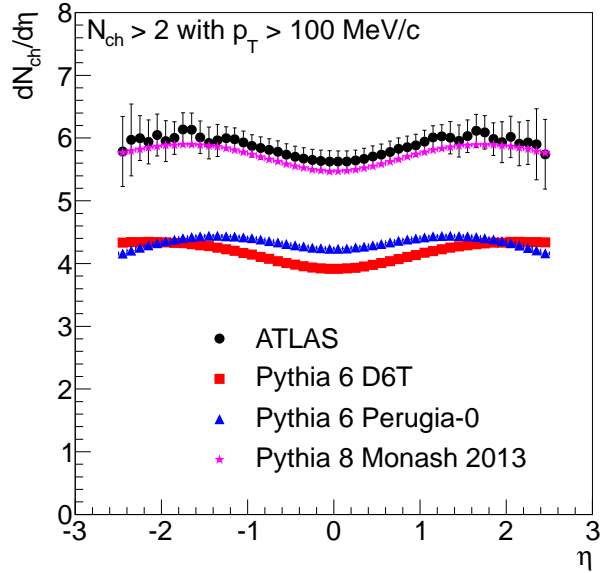}}
\caption{Comparison of $d\nch/d\eta$ at $\s = \unit[7]{TeV}$ (circles, from Ref.~\cite{Aad:2010ac}) with pre LHC tunes (Pythia 6~\cite{Sjostrand:2006za} D6T (squares) and Perugia-0~\cite{Skands:2010ak}  (triangles)) and Pythia 8~\cite{Sjostrand:2007gs} Monash 2013~\cite{Skands:2014pea} (stars) which has been tuned to LHC data. The unexpected increase in multiplicity is clearly visible by the large discrepancy between predictions and the data. The Monte Carlo simulation data are replotted from MCPLOTS~\cite{Karneyeu:2013aha}.}
\label{fig:mb_mcplots_dndeta}
\end{figure}

\begin{figure}[t!]
\centerline{\includegraphics[width=8cm]{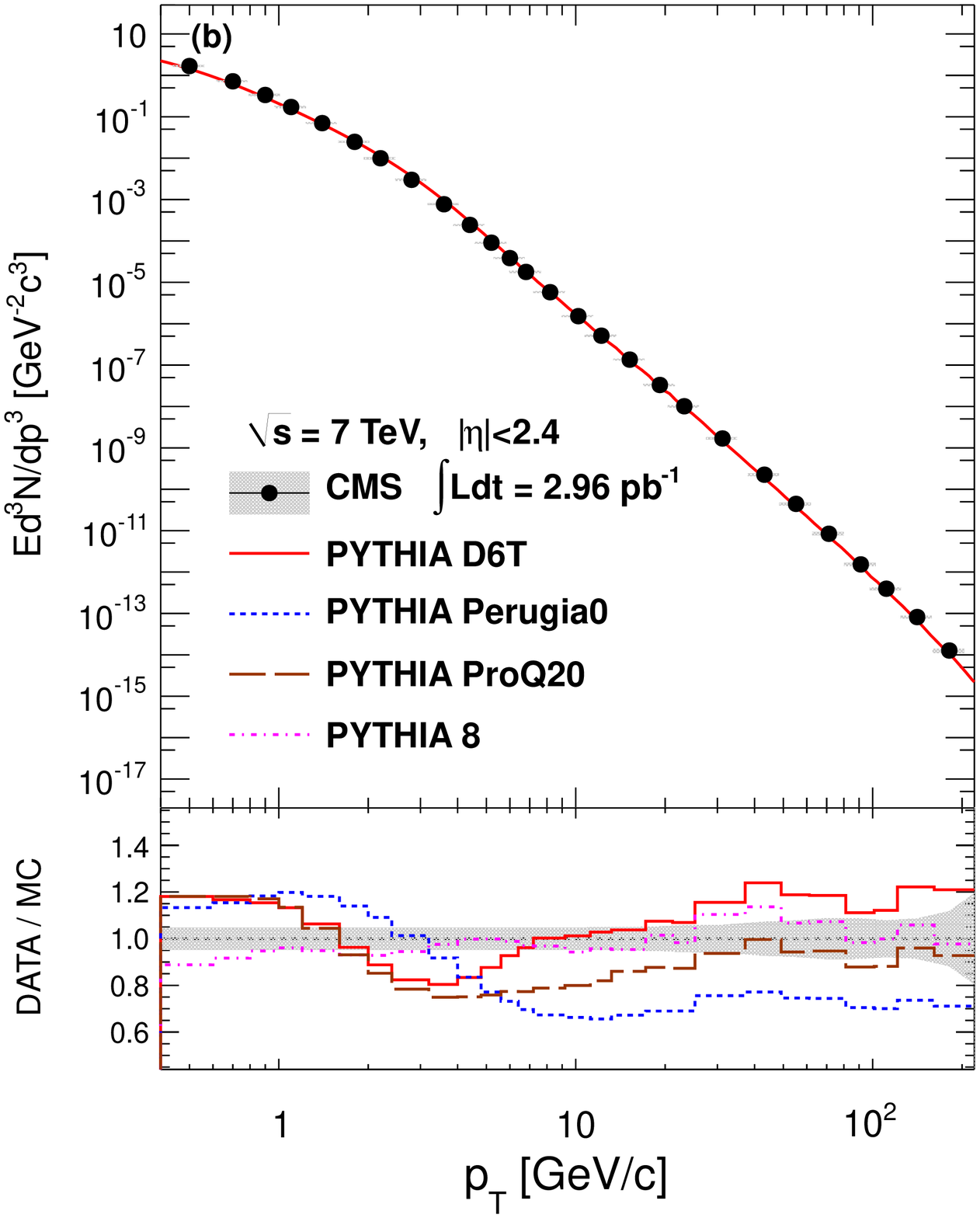}}
\caption{Transverse-momentum spectra up to \unit[200]{GeV/c} compared to several MC tunes which despite the impressive $y$ axis range over 14 orders of magnitude describe the data within 20\%. Figure from Ref.~\cite{Adam:2015pza}.}
\label{fig:mb_dndpt}
\end{figure}

The pseudorapidity density $d\nch/d\eta$ and the transverse-momentum spectra $d\nch/d\pt$ measure the average number of particles for a given event class. It is the most basic reduction of the complexity of a particle collision. 
Figure~\ref{fig:mb_dndeta} presents a compilation of $d\nch/d\eta$ from pre-LHC energies up to \unit[8]{TeV} at the LHC. With increasing $\s$, both, the height of the central plateau and the variance of the distribution grow. The dip around $\eta \approx 0$ is an artifact of the transformation from rapidity to pseudorapidity. In addition to the measurements around mid-rapidity by ALICE, ATLAS and CMS, LHCb and TOTEM have studied the forward region.

The growth of $d\nch/d\eta$ at $\eta = 0$ as a function of $\s$ is presented in Fig.~\ref{fig:mb_dndeta_vs_sqrts}. The dependence is described by a power-law as a function of $\s$ whose motivation is phenomenological.
The increase at LHC energies was unexpectedly large~\cite{Khachatryan:2010xs, Aamodt:2010pp} rendering many model predictions and tunes incorrect. This is interesting as the description of the increase of the average multiplicity in pQCD-inspired Monte Carlo models is sensitive to the $\s$ dependence of the lower momentum cut-off in the calculation of the $2 \rightarrow 2$ cross section and the proton matter distribution which both affect strongly the number of parton interactions occurring in the same collision (see Chapter~10 in Ref.~\cite{Bartalini:2017jkk}). The evolution of the MC tuning effort is illustrated in Fig.~\ref{fig:mb_mcplots_dndeta} which presents pre-LHC tunes and state-of-the-art Pythia tunes compared to early LHC results at \unit[7]{TeV}. While the former deviate from the data by 25--30\%, the latter accurately describes the data.

The transverse-momentum spectra $d\nch/d\pt$ combines the measurement of the soft regime at low $\pt$ with the hard regime at high $\pt$ which can be calculated in pQCD. While early measurements could focus only on the regime up to a few GeV/c, distributions have later been measured up to \unit[200]{GeV/c}~\cite{Chatrchyan:2011av, Aad:2015wga}. An example is shown in Fig.~\ref{fig:mb_dndpt}.

\subsection{Multiplicity distribution $P(\nch)$}

\begin{figure}[t]
\centerline{\includegraphics[width=8cm]{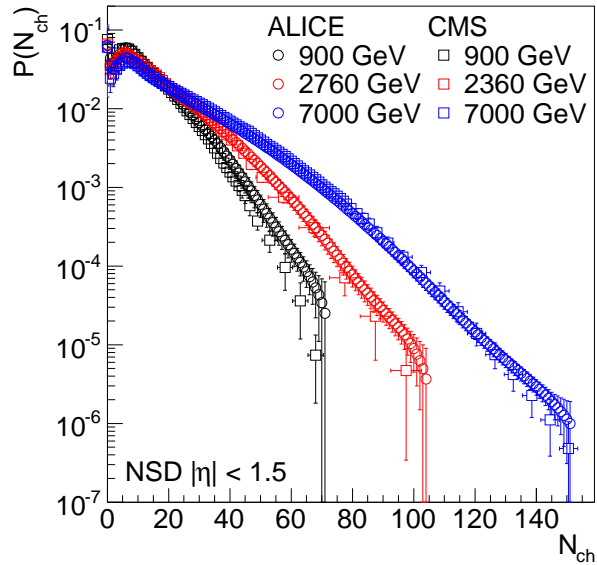}}
\caption{Multiplicity distribution in $|\eta| < 1.5$ for NSD collisions from \unit[900]{GeV} to \unit[7]{TeV} illustrating the wide tail of the multiplicity distribution at LHC energies. Data from Refs.~\cite{Adam:2015gka,Khachatryan:2010nk}.}
\label{fig:mb_mult}
\end{figure}

The multiplicity distribution $P(\nch)$ gives the probability that a collision has a certain number of charged particles. Due to the limited acceptance of the LHC experiments, the measurement is typically performed in a limited pseudorapidity range (contrary to pre-LHC experiments which often measured this distribution in full phase space). Figure~\ref{fig:mb_mult} presents a compilation of results from ALICE and CMS which measured $P(\nch)$ for NSD collisions in $|\eta| < 1.5$. In addition, to the good agreement between the two experiments, it can be seen that the width of the distribution grows significantly with $\s$ giving access to a very interesting high-multiplicity regime (see Chapter~15 in Ref.~\cite{Bartalini:2017jkk}). Recent measurements reach out to 250 particles in $|\eta| < 2.5$~\cite{Aaboud:2016itf}.

Historically, multiplicity distributions gave rise to a rich phenomenology~\cite{GrosseOetringhaus:2009kz}. KNO scaling~\cite{Koba:1972ng} asserted that distributions at all $\s$ fall onto a universal curve under a transformation dividing by the average number of particles. Negative binomial distributions (NBDs) were successful to describe $P(\nch)$ at SPS energies~\cite{Alner:1985zc} but failed at higher energies. Two-component approaches using two \cite{Fuglesang:1989st,Giovannini:1998zb} (or even three \cite{Giovannini:2003ft}) NBDs could not survive up to LHC energies (see e.g. Ref.~\cite{Adam:2015gka}). Nowadays, multiplicity distribution are a very sensitive probe of multiple parton interactions as collisions with large multiplicities are mostly composed of several parton interactions (see Section~\ref{sect:minijet}). Event generators fail to describe the tail of the multiplicity distribution without considering multiple parton interactions and the careful tuning of the related parameters.

\subsection{Mean transverse-momentum evolution}

\begin{figure}[t]
\centerline{\includegraphics[width=8cm]{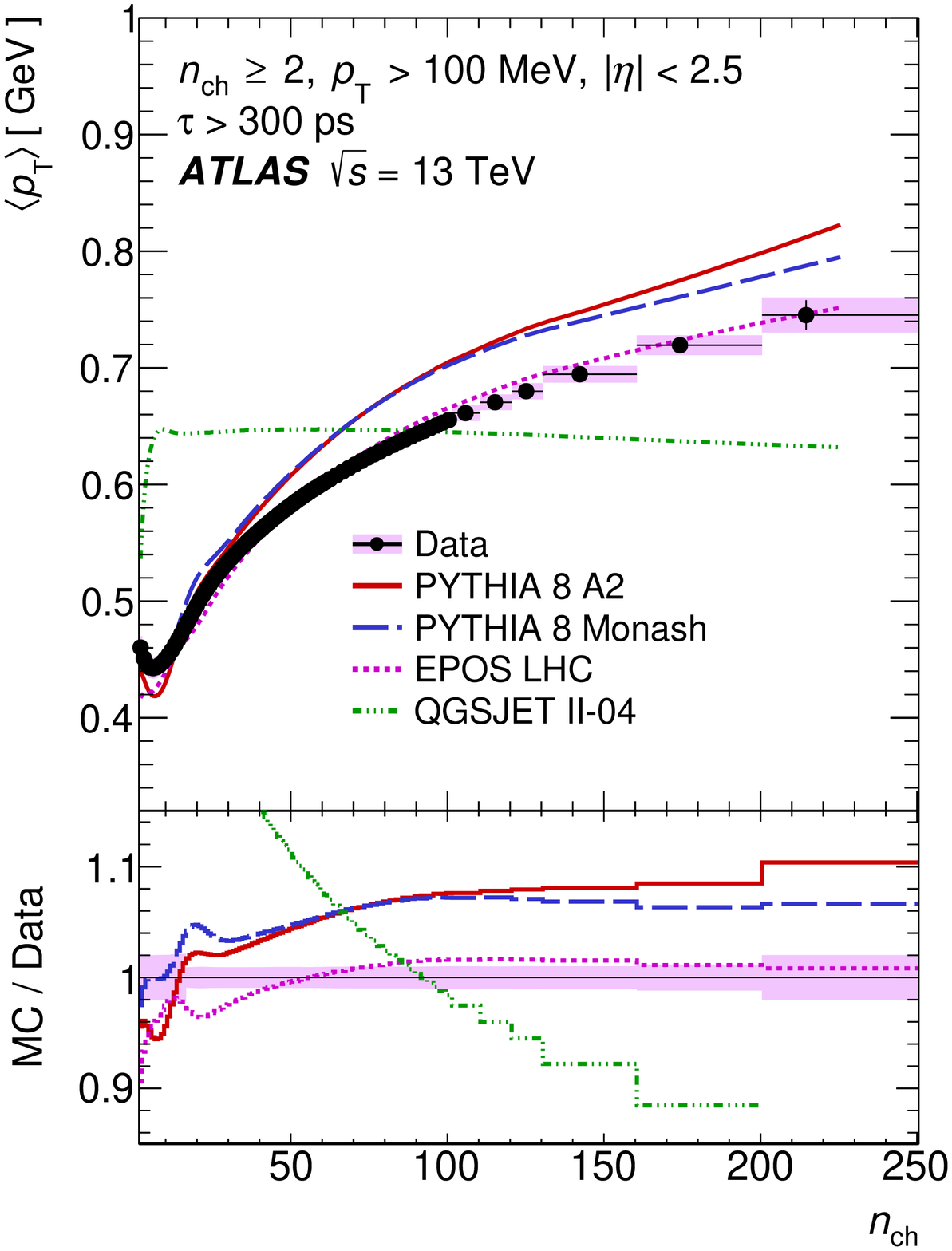}}
\caption{Average transverse-momentum $\meanpt$ as a function of $\nch$ in $|\eta| < 2.5$ at $\s = \unit[13]{TeV}$ compared to models. Figure from Ref.~\cite{Aaboud:2016itf}.}
\label{fig:mb_atlas_pt_vs_nch}
\end{figure}

The evolution of the mean transverse momentum $\meanpt$ as a function of $\nch$ measures the correlation of the momenta of the bulk of the produced particles with the multiplicity. 
It can differentiate if high-multiplicity events are \emph{simple} superpositions of low-multiplicity collisions or if coherent effects between different parton interactions have a significant influence.
Figure~\ref{fig:mb_atlas_pt_vs_nch} presents a recent result which demonstrates the increase of $\meanpt$ with growing $\nch$. State-of-the art Pythia tunes describe this observable within 10\% while the best description of this observable is provided by the EPOS LHC generator (see Chapter~19 in Ref.~\cite{Bartalini:2017jkk}). In the Pythia model, the correct description of this observable requires the color-reconnection mechanism (see Chapter~10).
Further insight of the growth of the $\meanpt$ is obtained using so-called underlying event observables where the activity is studied transversely to the hardest object in the event (see Chapter~11).

\subsection{Minijets} \label{sect:minijet}

\begin{figure}[t]
\centerline{\includegraphics[width=7cm]{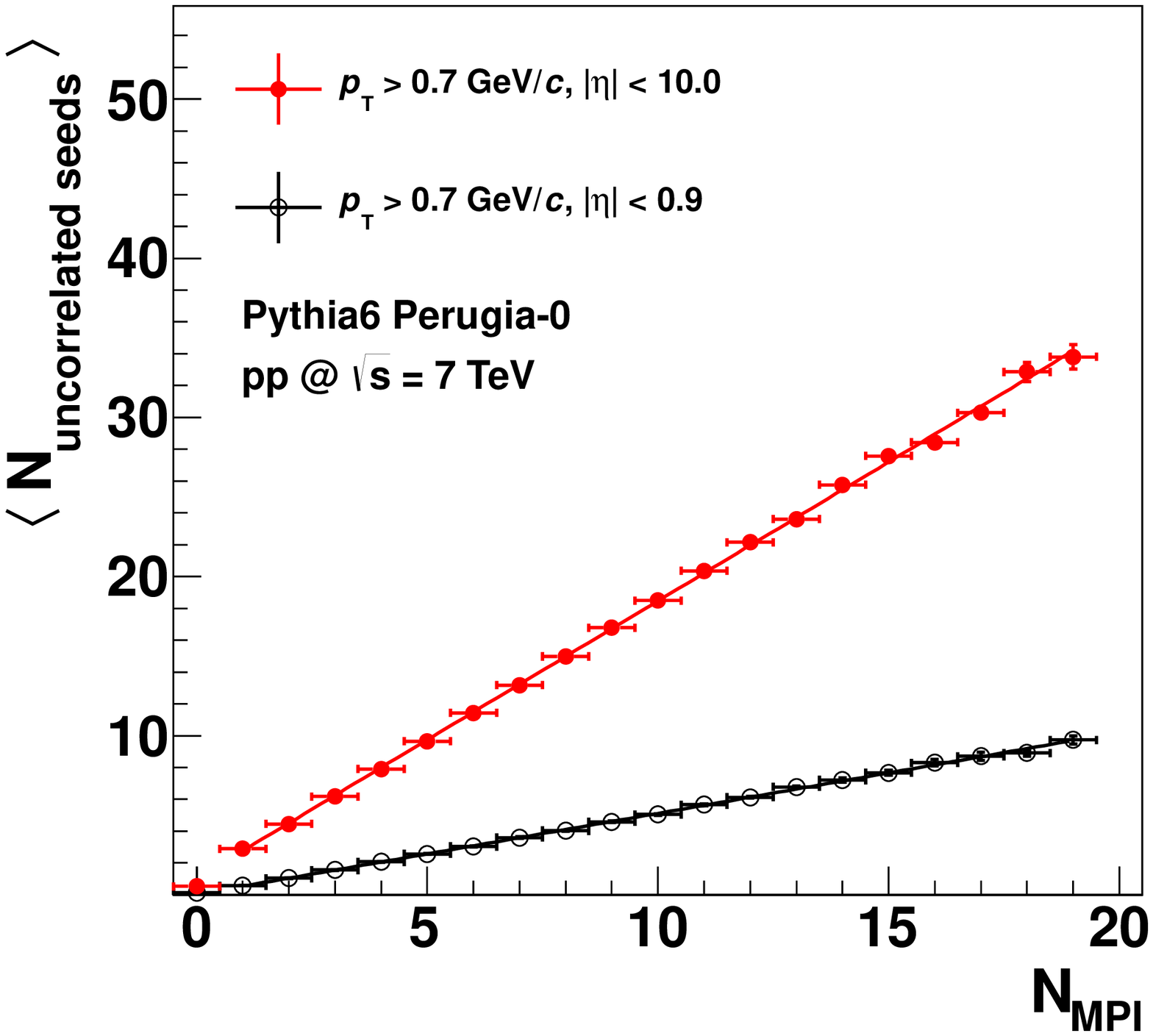}}
\caption{Correlation of the uncorrected seeds and the number of parton interactions showing the clear proportionality of these two measures within Pythia. Figure from Ref.~\cite{Abelev:2013sqa}.}
\label{fig:mb_minijet_pythia}
\end{figure}

\begin{figure}[t!]
\centerline{\includegraphics[width=0.49\columnwidth,clip=true,trim=0 0 40 50]{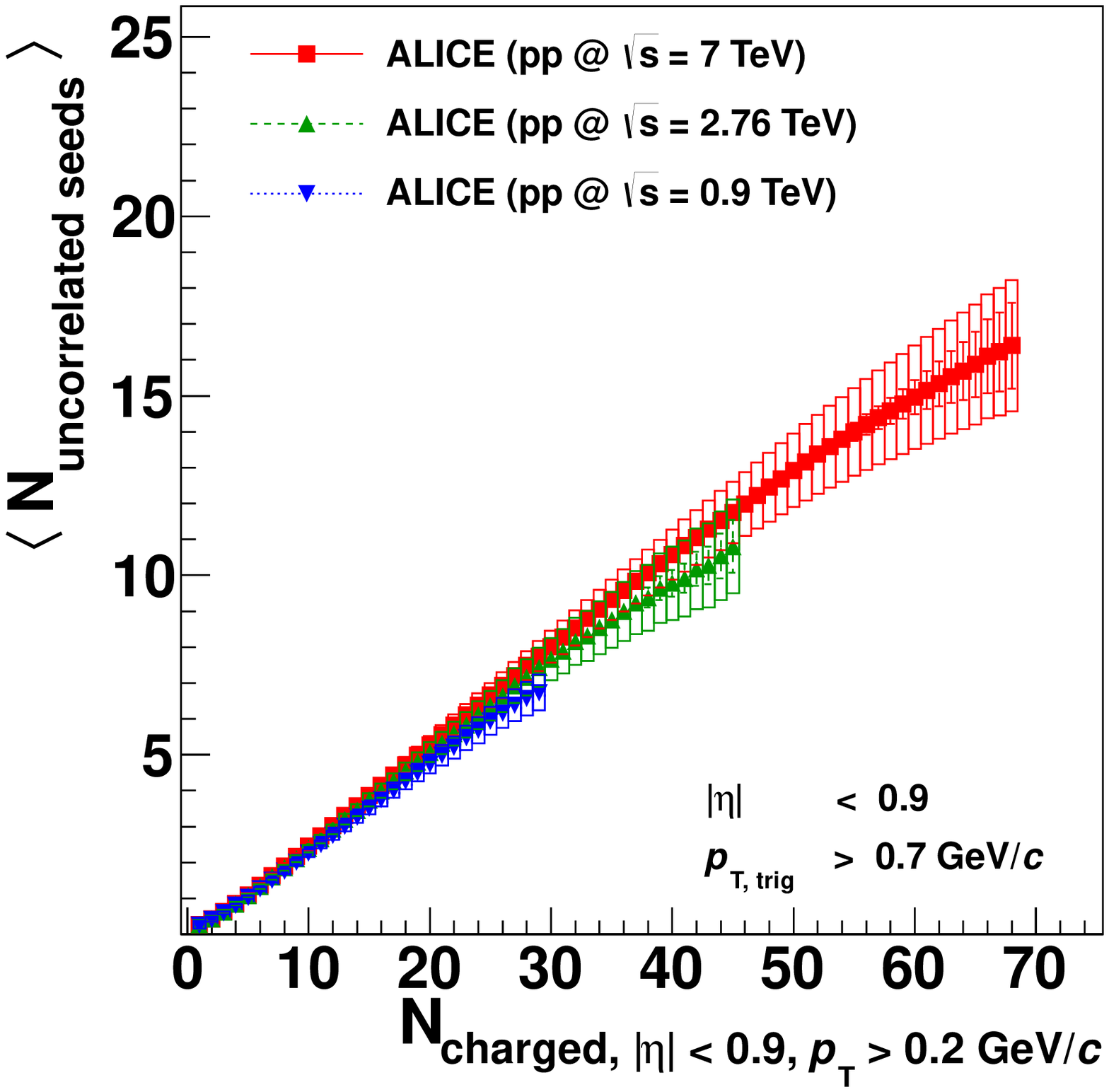} \hfill \includegraphics[width=0.49\columnwidth,clip=true,trim=40 0 0 50]{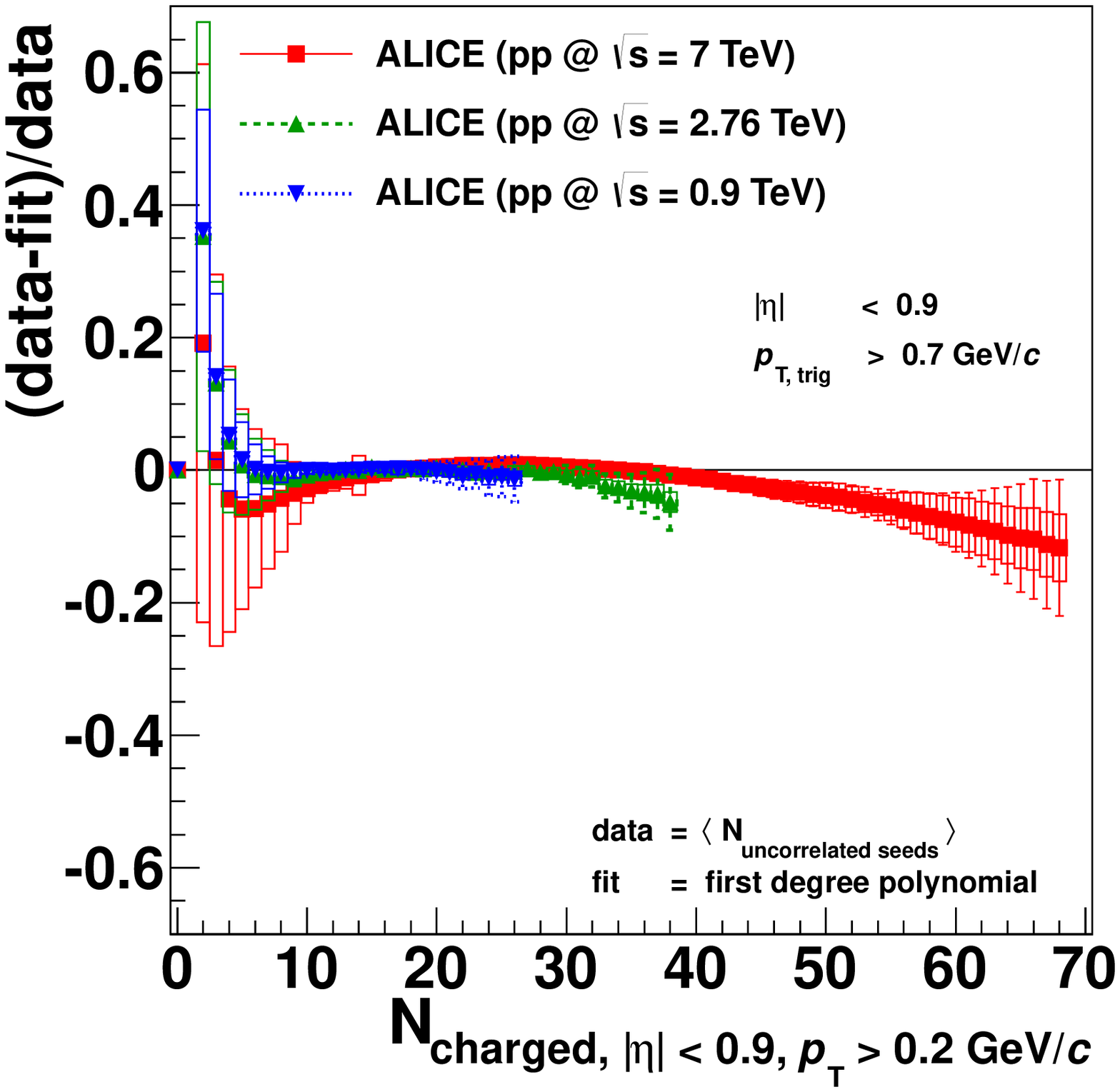}}
\caption{Number of uncorrelated seeds as a function of $\nch$ for $\s$ from \unit[900]{GeV} to 
\unit[7]{TeV} (left panel). The uncorrelated seeds increase almost linearly with \nch which is illustrated in the right panel where the difference to a linear fit is shown. At high $\nch$ a mild deviation from this linear trend is observed. Figures from Ref.~\cite{Abelev:2013sqa}.}
\label{fig:mb_minijet_us}
\end{figure}

The observables presented above give an indirect access to the number of parton interactions through MC tuning. A direct way is the measurement of so-called minijets~\cite{Abelev:2013sqa} which are constituted of particles stemming from the same $2 \rightarrow 2$ parton scattering, i.e. jets, but at $\pt$ of a few GeV/c where traditional jet reconstruction algorithms are not applicable. In this regime, each parton produces only 1--2 particles. Experimentally statistical methods like two-particle correlations allow nevertheless a measurement.
In this method, the number of associated particles to a so-called trigger particle are extracted on the near- and away-side. From this the uncorrelated seeds can be calculated which are a measure of the number of independent clusters in a collision~\cite{Abelev:2013sqa}: 
\begin{equation}
\langle N_{\rm uncorrelated\ seeds} \rangle = \frac{\langle N_{\rm trigger} \rangle}{\langle 1 + N_{\rm assoc,near-side} + N_{\rm assoc,away-side}}.
\end{equation}
Figure~\ref{fig:mb_minijet_pythia} illustrates that this quantity is proportional to the number of parton interactions in Pythia.
Figure~\ref{fig:mb_minijet_us} shows the measured uncorrelated seeds for $\s = \unit[0.9]{TeV}$ to \unit[7]{TeV} as a function of $\nch$. The results at different energies are very similar at fixed $\nch$ (left panel). In addition at large multiplicities a deviation from a linear trend is visible (shown in the right panel) hinting at a saturation in the number of parton interactions. It would be very interesting to extend this measurement to higher \nch and study if indeed above a certain \nch no further increase of the number of parton interactions is observed.

\section{Outlook}

This chapter has presented selected results of minimum-bias results at the LHC, $d\nch/d\eta$, $d\nch/d\pt$, $P(\nch)$ and $\meanpt$ vs. $\nch$. In addition, the minijet observable was introduced which gives direct access to the number of parton interactions. 
These measurements constitute a precise characterisation of the average LHC collision. At the same time, they study rare collisions at high multiplicity. Future LHC data-taking will allow to extend this into the regime of very large multiplicity and very large number of parton interactions -- a region full of interesting aspects of soft QCD.

\bibliographystyle{utphys}
\bibliography{biblio,reviewbib}

\end{document}

%% file: commands.tex
\newcommand{\s}            {\ensuremath{\sqrt{s}}}
\newcommand{\meanpt}       {\ensuremath{\langle p_{\mathrm{T}}\rangle}{ }}
\newcommand{\pt}           {\ensuremath{p_{\mathrm{T}}}{ }}
\newcommand{\nch}          {\ensuremath{N_{\mathrm{ch}}}{ }}

\newcommand{\com}[1]       {}